# Empirically-Calibrated H100 Node Power Models for Reducing Uncertainty in AI Training Energy Estimation

Alex C. Newkirk[1,2], Jared Fernandez[3], Jonathan Koomey[4], Imran Latif[5], Emma Strubell[3], Arman Shehabi[6], Constantine Samaras[2,7,8]

## Abstract

As AI's energy demand continues to grow, it is critical to enhance the understanding of characteristics of this demand, to improve grid infrastructure planning and environmental assessment. By combining empirical measurements from Brookhaven National Laboratory during AI training on 8-GPU H100 systems with open-source benchmarking data, we develop statistical models relating computational intensity to node-level power consumption. We measure the gap between manufacturer-rated thermal design power (TDP) and actual power demand during AI training. Our analysis reveals that even computationally intensive workloads operate at only 76% of the 10.2 kW TDP rating. Our architecture-specific model, calibrated to floating-point operations, predicts energy consumption with 11.4% mean absolute percentage error, significantly outperforming TDP-based approaches (27-37% error). We identified distinct power signatures between transformer and CNN architectures, with transformers showing characteristic fluctuations that may impact grid stability.

[1] Building Technologies and Urban Systems, Energy Technologies Area, Lawrence Berkeley National Laboratory, Berkeley, CA, USA

[2] Department of Engineering and Public Policy, Carnegie Mellon University, Pittsburgh, PA, USA

[3] School of Computer Science, Carnegie Mellon University, Pittsburgh, PA, USA

[4] Koomey Analytics, San Francisco, CA, USA

[5] Computing and Data Sciences Directorate, Brookhaven National Laboratory, Upton, NY, USA

[6] Energy Analysis and Environmental Impacts Division, Energy Technologies Area, Lawrence Berkeley National Laboratory, Berkeley, CA, USA

[7] Department of Civil and Environmental Engineering, Carnegie Mellon University, Pittsburgh, PA, USA

[8] Wilton E. Scott Institute for Energy Innovation, Carnegie Mellon University, Pittsburgh, PA, USA

# Introduction

## Background and Motivation

The increased capabilities of artificial intelligence (AI) applications have led to their rapid integration into business and consumer ecosystems. This demand has been accompanied by large-scale investment in the data centers that constitute the physical substrate of these services [1]. Properly anticipating the significant power and transmission infrastructure needs of future data center facilities, as well as assessing their growing environmental impact, depends on accurately measuring hardware power consumption and understanding how AI computation differs from traditional enterprise workloads.

Excitement around the economic potential of AI has led to a proliferation of AI in common parlance and marketing materials. AI can refer to any applications where analytic insights, novel images, or text are generated through computer algorithms, as well as image classification, natural language processing, and advanced automation. When we refer to AI, we are specifically discussing the computationally intensive applications built on machine learning (ML) techniques. Innovations of the last decade such as deep-neural-nets (DNN) [2] and the transformer architecture [3] have enabled both greater performance in AI models and greater computational efficiency on parallel hardware. These advancements repeatedly illustrated "the bitter lesson" of AI research: the most dramatic performance improvements have been seen in "methods that continue to scale with increased computation, even as the available computation becomes very great" [4] including deep learning techniques. This scaling relationship between computation and capability has profound implications for facility infrastructure, as improvements in model performance have consistently been reinvested in larger models with greater computational demands.

Modern AI systems are composed from three fundamental inputs: datasets that contain information about our world, algorithms that process that information, and physical computer infrastructure to implement the data processing algorithms [5]. Training data, which can vary from physical modeling parameters to natural language to images, are vectorized [6] (converted into a column of numeric values). This vector object can then be transformed using matrix operations. This is the mechanical core of DNN AI, the weightings and matrix operations are derived to best predict an output given some input data. A relevant feature of this computation is its floating-point simplicity compared to conventional high-performance computing (HPC) workloads; while each individual operation is relatively straightforward, a tremendous quantity of these operations must be performed.

Initially the model has either no weighting or random weightings, developing the final relationships between input and output data entails feeding the training set through the model, evaluating the predictive performance, adjusting the parameters, and repeating. Once developers have built an optimized model, they can then feed it queries and receive outputs based on its internal weightings. The process of developing the model weights is known as "training" the model, while querying a completed model and receiving output is known as "inference". The key insight from this is that AI applications require large quantities of mathematical operations that are not individually complicated, which makes them uniquely suited to parallel computation hardware.

Transformer architecture DNNs benefit from a variety of parallelization techniques including data parallelism, pipeline parallelism, and tensor parallelism[1] [7], which unlock increased computational

---

[1] Tensors are mathematical objects that satisfy certain transformation relationships conceptually similar to vectors and matrices. Tensors are commonly used in physics and engineering, and computing their time evolution requires

performance. Logic chips optimized for AI related workloads are sometimes collectively referred to as AI accelerators/accelerated processors or "AI Chips" [8]. The most common type of accelerator is the graphics-processing-unit (GPU), but there are a variety of accelerator architectures. For a more detailed discussion of the AI hardware stack, see Appendix A. Each parallelism technique affects hardware utilization differently: tensor parallelism shards the matrix operations for an individual layer across devices, pipeline parallelism distributes model layers across devices, while context parallelism (and sequence parallelism) divides attention operations across GPUs to reduce memory requirements for processing long sequences [9].

The bulk of the computational muscle supporting AI is delivered by AI accelerators. Groups of AI chips are typically integrated with one of more central-processing-units (CPU) to coordinate workloads, supervise data transfer, and interface with other systems, as well as significant memory systems, and high-bandwidth networking components. This bundle of hardware is often referred to as a compute node, a term analogous to server, with a group of nodes referred to as a cluster.

Electricity use in data centers is largely driven by the servers, data storage, and networking equipment they house. Computation generates resistive heat, which must be removed from the data center to maintain equipment reliability and longevity. The total electricity use of the facility is directly related to the electricity used by the information technology (IT) equipment, which is why accurately assessing IT electricity use in these facilities is critical. Analyses of total data center electricity use have traditionally characterized the non-IT electricity use in data centers using a ratio called power usage effectiveness (PUE).

PUE is defined as the total electricity use of the facility divided by the IT electricity use of the facility measured at the plug. It is a ratio that is always larger than 1.0, and the fraction above 1.0 represents the losses and electricity consumed by external fans, pumps, power distribution, and cooling (generally characterized as "infrastructure electricity use"). The form of this ratio embodies the primacy of IT electricity use in the total electricity used by the facility. If IT electricity use goes up, the infrastructure electricity use goes up in direct proportion, all other things being equal. Errors made in assessing IT electricity use therefore compound across any model or measure of facility energy use.

Our analysis focuses specifically on IT electricity use for several reasons. First, operating electricity has historically dominated data center environmental footprint, accounting for over 90% of lifecycle emissions [10], although this share will continue to decline as grids decarbonize [11], [12], [13], [14]. While estimates of IT hardware embodied emissions have significant uncertainty [15], [16], [17], previous analyses of enterprise computing equipment found embodied carbon accounts for only 2-8% of lifecycle emissions [18]. This operational dominance is likely applicable for AI hardware given its high utilization and power density compared to conventional servers. Life-cycle assessment of tensor processing units estimated that operational emissions accounted for 90% of total emissions in the baseline case, dropping to 70% when accounting for zero-carbon power purchase agreements [19].

Second, facility siting, and by extension the carbon and water intensity of electricity supply [20] and the feasibility of free cooling, is the primary determinant of a data center's overall environmental footprint [21]. As infrastructure energy use scales directly with IT load (i.e., cooling load scales with the waste heat generated by the IT hardware), accurate assessment of IT electricity use is fundamental to understanding

---

matrix operations akin to machine learning. Accordingly, the term tensor is sometimes shorthand used to describe operations of this kind and to signify hardware optimized for these applications, e.g. TensorFlow, Tensor-Processing-Unit, etc.

both total facility energy consumption and environmental impact. Accordingly, this work prioritizes developing robust estimates of IT power demand during AI training.

## Estimating AI Electricity Use

Historically, most assessments of AI related energy use have focused on the training of AI models. This in part stems from the distinctive profile of the physical hardware: training occurs on specialized, heat-dense processors in hyperscale facilities. While these scale and finite boundary characteristics are often perceived as making it straightforward to model and attribute energy use to the training of a particular model, in reality this process is considerably more complex and challenging than commonly understood [22]. It also reflects industry norms; hyperscale cloud providers typically bill ML training customers according to the number of accelerated chips they provide and their hours of use. Thus, GPU-hours serves as industry shorthand for model scale alongside parameter count.

One approach to modeling AI training energy use is based on empirically measured power draw of key hardware components [23], [24], [25], [26] multiplied either by total mathematical operations [27] or by GPU-hours [28], [29]. As models have increased in size, so too have the upfront capital costs of training [30], and the ability to operate these facilities efficiently has become a key source of competitive advantage with publicly available information becoming scarce. Methodologically rigorous estimates of AI energy-use often rely on proprietary power draw data [11], [31], [32].

In practice, many estimates of AI training power consumption multiply the manufacturer rated maximum power draw (also referred to as *thermal design power* or TDP) of AI chips [33] or servers [34] by the reported training time (typically in GPU-hours) of a given model [35], [36]. This reflects data availability, as real time power measurements are not typically retrievable months or years after training has occurred [28], and multiplying GPU-hours by chip TDP is a straightforward modeling approach. These models will then multiply this output power by a parameterized PUE to arrive at a total energy footprint. These studies typically link their power estimates to a spatial-temporal analysis of carbon emissions [37]. Meta estimates the carbon footprint of training their LLAMA models through this chip TDP and GPU-hours approximation [38], [39]. Calibrating these estimates with empirical grounding serves to shed further light on the energy footprint of AI applications, and improve model accuracy [40]. While performance benchmarks have been proposed and implemented in the industry [41], training benchmarks have not kept pace with commercial system scale up [42].

Scholars typically either report the individual chip level energy performance [43], [44], or aggregate facility level power draw [33]. Data center energy use model validation requires chip, node, and system-level power measurement [24], and while GPU-level energy performance is well specified [31], nodes and clusters are not. Reliable documentation of node-level hardware performance for current generation workloads and servers is scarce, meaning even knowledgable industry actors having to rely on rules-of-thumb in their estimates of in-use node power draw [45]. Our work addresses this gap for current generation hardware by developing an empirically calibrated statistical model of H100 node power draw during training. While this work primarily focuses on the energy footprint of training, see Appendix B for an overview of inference energy use and associated literature.

This paper makes the following primary contributions to the field: We analyze an integrated dataset of empirical measurements of AI training power demand on a single node combined with open-source training data on comparable workloads across single and multi-nodal training configurations. Through this analysis, we find that actual power draw consistently remains well below thermal design TDP, with even computationally intensive workloads drawing on average no more than 76% of TDP. We then use these data to develop an empirically calibrated statistical model of 8-GPU H100 DGX node-level power draw based on FLOPs per node, which achieves 11.5% mean absolute percentage error in predicting training workload power draw across our test workloads - significantly outperforming TDP-based approaches which result in errors of 27-37%. Our analysis revealed distinct power signatures between CNN and transformer architectures and characterized how node-level power demand varies with training configuration. Finally, we discuss these findings in the context of the energy system, identifying key drivers of uncertainty in future efforts to assess the energy footprint of AI computation.

## Methods

Our empirically calibrated statistical analysis relies on a dataset constructed by combining two complementary sources of AI training power measurements. The first consists of detailed empirical measurements collected at Brookhaven National Laboratory (BNL), which include time-series data of node-level power demand on an 8 NVIDIA H-100 DGX during AI training benchmarks, as well as node idle power and GPU stress test data. Complete documentation of these measurements and the raw time-series data are available in Latif et al. [46]. Those measurements were collected on a single-node, and so may not generalize to the multi-nodal training parallelization which defines the industry [47], [48], where data communication introduces additional complexity [49]. To address this limitation, we integrated the BNL data with open-source multi-node training data for comparable workloads, creating a dataset that spans both single-node and distributed training environments. Using these data, we developed a novel statistical model of node-level average power demand during AI training.

Training power draw data were collected from *MLPerf® Training v4.0* [50] power measurement submission by *Sustainable Metal Cloud* (SMC)[2]. Initial data were retrieved on August 12th, 2024, from run IDs 4.0-0090 through 4.0-0096 (excluding 4.0-0093). We selected these open-source data due to the mixture of workloads and training configurations. We only analyzed the training runs which had a direct workload proxy to the BNL data, namely image classification (Resnet) and NLP (GPT3-175 and Llama). Training workloads without analogues in the BNL data we included in the validation set.

The hardware configuration of the SMC nodes does differ from the node at BNL. Both are dedicated DGX accelerated AI training servers with 8 NVIDIA H100-SXM5-80 GB GPUs. The manufacturer rated TDP for these systems is 10.2 kW [51]. SMC nodes employ two *Intel Xeon Platinum 8462-Y* model processors as supervisory CPUs, and are specified as containing 2,000 GB of memory per server. In contrast, the BNL model contains 1,500 gigabytes of memory and the supervisory CPUs are *AMD EPYC 9354.* Additionally, the SMC nodes are connected by NVIDIA specified *Infiniband* interconnect fabric. These are the differences available in publicly available documentation, without detailed bills-of-material we can only approximate differences between the hardware. Additionally, SMC nodes recorded power measurements approximately every two seconds, compared to the 5-minute interval for the BNL data. While non-trivial, we can account for these differences through our choice of statistical techniques.

When error terms are correlated within clusters but independent across clusters, then conventional standard errors, which assume independence between all observations, will contain statistical bias [52]. We know our data contains some of these confounding factors, such as the differences in hardware configurations. Cluster-robust standard errors are designed to allow for correlation between observations within cluster [53]. Additionally, an unweighted regression of our data would in practice weigh the estimator towards the workloads with the fastest measurement interval and most nodes. To address these issues, we performed a weighted regression with robust clustered standard errors. Note that clustered standard errors do not affect the estimated parameter, just the estimation of the errors. We will now discuss our model specification and interpretation in greater detail.

---

[2] MLPerf® Training v4.0 Power measurement benchmark. Retrieved from https://github.com/mlcommons/training_results_v4.0/tree/main/smc/results on August 12, 2024, entries 4.0-0090, 4.0-0091, 4.0-0092, 4.0-0094, 4.0-0095, 4.0-0096. The MLPerf name and logo are registered and unregistered trademarks of MLCommons Association in the United States and other countries. All rights reserved. Unauthorized use strictly prohibited. See www.mlcommons.org for more information.

## Model Specification

Our objective was to develop a statistical model that could predict node-level power-draw across different AI training workloads and system configurations.. To develop this model, we began by considering the physical structure of power consumption in an AI training server. For a more detailed discussion of AI specialized IT hardware and facility characteristics see appendix A.

A server consists of several components that contribute to total power consumption. These components include logic elements that perform computations, memory systems that store data and intermediate results, and interconnect fabric that enables communication between them. When training an AI model, the primary computation occurs in the parallel logic hardware, in our case the GPUs, which account for the largest share of power demand in a node [54]. Note that NVIDIA specified DGX servers contain 12 fan modules "behind the meter". While not strictly IT load in the ideal theoretical definition PUE, for the purposes of metering the power supplied to a server or rack, this demand is bundled with the computational components.

At the most basic level, a server's total power demand (*Y*) can be understood as the sum of power consumed by its constituent components:

$$Y = \sum_{i \in C} P_i \tag{1}$$

Let set *C* represent the power-consuming components in a server: GPU accelerators (*G*), supervisory logic (*U*), memory and storage systems (*D*), interconnect (*I*), and cooling fans (*F*).

$$C = \{G, U, D, I, F\} \tag{2}$$

While each component in set *C* could theoretically be measured independently, our data records aggregated power measured at the node level. The total power demand of computational hardware fundamentally scales as some function of the number of operations it performs (x) [55], [56]. For each component *i* in set *C*, we can express this relationship as:

$$P_i = \phi_i(x) \tag{3}$$

Logic chips, as the physical substrate performing the operations, have the most intuitive connection between operations and power draw, though the practical scaling relationship demonstrates complexity[57]. The other components in a server[3] can have more complex relationships with computational load. In some cases, such as embedded fans, these relationships are theoretically straightforward: cooling load scales with heat generation, which results from resistive losses in electrical components, creating a linear relationship between single-speed fan power and the electricity consumed by other components (assuming no external cooling infrastructure). This is a first order approximation, for a more detailed discussion of fan power, see appendix C.

---

[3] In the case of an AI server this includes the supervisory logic devices which coordinate different server elements but do not themselves conduct the training computations

The power demands of memory and interconnect components have more complex scaling behavior. These components both enable and respond to logical operations - memory must store inputs and intermediate results, while interconnect facilitates data transfer between components. This means they may scale independently with total operations, they may be dependent on logic chip computation, or they may be a mixture of direct or indirect dependencies. While future work could illuminate these complexities through component-level submetering, for our purposes modeling aggregate node power as a function of computational intensity captures these interactions while remaining tractable.

Therefore, the total power supplied to a server can be expressed as:

$$Y = \sum_{i \in C} \phi_i(x) \qquad (4)$$

A finite piece of hardware can only perform some maximum quantity of operations within a given time [58]; more prosaically there would be some physical limit of computations where a processor would generate more heat than cooling systems could dissipate, causing thermal runaway and destroying the hardware. While individual components may scale differently with computational load - some linearly up to a hard limit, others asymptotically approaching a maximum - all must eventually reach some finite maximum power. Therefore, for any component *i*:

$$P_i \leq P_{i,max} \qquad (5)$$

The exact functional form of $\phi_i(x)$ may vary by component type [57], see Appendix C for a more detailed discussion, but the aggregate node power demand will be bounded by the sum of these component maxima at computational saturation, a quantity we define as *(P{comp})*:

$$Y \leq \sum_{i \in C} P_{i,max} \qquad (6)$$

$$P_{comp} = \lim_{x \to \infty} \sum_{i \in C} \phi_i(x) = \sum_{i \in C} P_{i,max} \qquad (7)$$

There are several candidate functional forms which could describe this system. The simplest specification assumes the server begins at some idle power ($P_{idle}$) at zero computational demand, and then approaches some asymptotic limit as ($P_{max}$) as computational demand increases. The difference between $P_{max}$ and $P_{idle}$ is then $P_{comp}$, with the parameter $\alpha$ governing the steepness of the transition:

$$Y = P_{idle} + P_{comp} \frac{x}{\alpha + x} \qquad (8)$$

We hypothesize that the actual power behavior follows a sigmoid function, as illustrated by equation 9. In this formulation, total power (Y) depends on the baseline power draw of an idle server (*P{idle}*), the maximum additional power draw from active components (*P{comp}*), the quantity of operations being performed (*x*), an operations threshold (*x{o}*) governing the midpoint of demand saturation and a parameter (*k*) controlling the steepness of transition at which active power approaches the saturation value. Y, $P_{idle}$, and *P{comp}* would be in units of Watts, *x* and *o* would be in units of operations, and *k* would be dimensionless.

$$Y = P_{idle} + P_{comp} \frac{e^{(x-x_0)/k}}{1 + e^{(x-x_0)/k}} \qquad (9)$$

This specification captures several important physical characteristics of the system. The offset parameter $x_0$ represents a threshold of computational operations that can be performed using hardware resources already active for system maintenance and monitoring. Below this threshold, the marginal power demand for additional operations is minimal, as these calculations can be performed by otherwise underutilized cores which have been activated to perform for basic system functions. The parameter k determines how sharply power demand increases once operations exceed this threshold, reflecting how quickly the system transitions from idle to active states.

While the sigmoid model might better represent the underlying physics of the system, particularly at low computational loads, it requires estimating additional parameters that present challenges with our limited empirical dataset. Each additional parameter introduces greater uncertainty and increases overfitting risk, especially when working with measurements across heterogeneous hardware configurations. The multicollinearity between transition parameters (k and $x_0$) further complicates estimation, as different combinations can produce similar curve shapes, creating ridge issues in the optimization landscape.

Given these constraints, we prioritize robust estimation of the asymptotic power limit ($P_{max}$), which directly informs infrastructure planning and energy forecasting. Since both functional forms converge to similar asymptotic limits as computational demand saturates, their practical difference lies primarily in how they model the transition from idle to active states. To determine which specification better serves our purpose, we evaluate both the asymptotic and sigmoid models using out-of-sample prediction accuracy as our primary metric, focusing specifically on each model's ability to generalize across different AI architectures and training configurations. This approach emphasizes practical utility for energy forecasting rather than theoretical elegance. We must choose some value for computational intensity. While the count of total model parameters provides a consistent measure of model scale across architectures, it represents an incomplete proxy for computational intensity driving node-level power demand. The relationship between parameters and actual computation is highly complex - different architectures require varying operations per parameter, hyperparameter choices impact computational demands independently, and the distribution of computation across hardware varies significantly based on parallelization techniques (data, tensor, and pipeline parallelism). Parameters fail to capture critical factors like activation functions, memory access patterns, and communication overhead between processing units.

Another potential proxy for computational demand is the expected required floating-point operations (*M*) to train a given model. Unlike parameter count, which insufficiently captures the computational complexity of different model architectures, FLOPS directly represents the arithmetic operations required during training, providing a more precise measure of computational demand. For transformer architecture models, we adopted the established methodology from Narayanan et al. (2021), which accounts for the specific computation patterns in these architectures, as shown in equation 10:

$$M_{LLM} = 96 \times B_{global} \times S \times L \times H^2 \times \left(1 + \frac{S}{6H} + \frac{V}{16LH}\right) \qquad (10)$$

Where $B_{global}$ is the global batch size, *S* is sequence length, *L* is the number of layers, *H* is hidden dimension size, and *V* is vocabulary size. The coefficient 96 represents six operations per non-embedding parameter in both forward and backward passes (2 × 3 × 16), as empirically determined by Narayanan et al. (2021). This formulation accounts for the attention mechanisms unique to transformer architectures and the computational patterns of both forward propagation and backpropagation. Note as is documented in the configuration files, global batch sizes for the language models were adjusted according to parallelization hyperparameters. Global batch size ($B_{global}$) was selected according to the relationship defined in equation 11:

$$B_{global} = B_{mini} \times \frac{Total\ GPUs}{TP \times CP \times PP} \quad (11)$$

Where $B_{mini}$ is the minibatch size, *TP*, *CP*, and *PP* are tensor, context, and pipeline parallelism respectively. These parallelization strategies distribute computation across hardware resources in fundamentally different ways: tensor parallelism (TP) splits individual operations across devices, context parallelism (CP) processes different sequences independently, and pipeline parallelism (PP) assigns different network layers to separate devices. Each strategy affects both computational efficiency and power distribution across the training cluster. Note that the benchmark submission LLM workloads adjusted global batch size based on their parallelization choices, creating an inherent interaction that cannot be isolated. While this collinearity prevents modeling each parallelization strategy's independent effect, our FLOPS-based computational proxy incorporates these factors through their collective impact on the derived global batch size calculation.

For convolutional neural networks (CNNs), we calculated FLOPS based on the known computational requirements of specific architectures, adjusted for image dimensions and accounting for both forward and backward passes [59], as defined in equation 13:

$$M_{CNN} = 3 \times M_{image} \times \left(\frac{I_{train}}{224}\right)^2 \times B_{global} \quad (12)$$

Where $M_{image}$ is the required flops per image per forward pass for a particular model for a standard 224×224 input, $I_{train}$ is the sidelength of the image in the training set, and $B_{global}$ is the global batch size. Image size was sourced from the workload documentation, and $M_{image}$ was sourced from the research literature [60]. We then divided these estimates of the required operations for an individual iteration of a given model by the number of nodes to get the effective flops per node in a given workload ($M_{eff}$), which we use as our computational proxy (*x*). This computational intensity proxy enables meaningful comparison across diverse model architectures and training configurations, capturing the interaction between model complexity, batch size, and parallelization strategies that directly influence node-level power demand.

All architectural parameters (hidden dimensions, layer counts, etc.) were sourced directly from the relevant model cards or benchmark workload documentation, while training hyperparameters (learning rate, batch size, etc.) were extracted from the benchmark configuration files. It is also noteworthy that effective FLOPS may not be ideal, as the energy intensity of operations can vary at a macro-system [40] and chip-level operational [61] level. A summary table of the architectural and hyperparameter values used for each workload is available in Appendix D. A review of the literature on empirical scaling laws suggested that computational demand scales with the log of computational intensity [62], [63], [64]. This is the underlying physical model we statistically approximate: average node-level power draw ($\hat{Y}$) approaches an asymptotic limit, the sum of idle power and estimated active power ($\beta_{comp}$), as the log of computational intensity (*x*) increases. This asymptotic model specification is shown in equation 13:

$$\hat{Y} = P_{idle} + \beta_{comp} \frac{log(x)}{\alpha + log(x)} \quad (13)$$

Note that this is an estimate of IT power, and does not include power switching losses, cooling load, or any other infrastructure energy use. While these are key considerations for assessing the environmental footprint of a given data center or AI model, they scale with IT demand [65]. We compared these IT power estimates to the sigmoid specification in equation 9.

Each of the nine workloads differed in the time required to complete training. To avoid overfitting on longer, more frequently sampled, and higher-node count training runs, we weighted according to workload in our regression. The SMC data were sampled every 2 seconds while BNL measurements were taken at 5-minute intervals, creating potential bias. Rather than discard data through downsampling to standardize measurement frequency, we preserved all observations and addressed these sampling biases through proportionate weighting scheme. We applied a weighting factor of 1/n to each observation within a workload, where n represents the total number of observations for that particular workload, ensuring that the cumulative weight of each workload summed to 1 regardless of its sampling frequency or duration By weighting according to workload rather than individual measurements, we ensure each training run contributes equally to our parameter estimates regardless of its duration, sampling frequency, or number of nodes. To evaluate the cluster robustness of our coefficients and standard errors, we clustered by individual workload.

We took the raw .csv and .txt time series of the power draw for each node training each of our workloads, and read them into statistical software, giving each individual node a unique identifier. Measurements were time-tamped according to time since training began. Interconnect power use in multi-nodal systems was allocated by dividing the total interconnect power use for each workload reported in the MLPerf submission by the number of nodes conducting training. For the transformer architecture models (GPT-3 and Llama), we sourced parallelism factors and batch size data from the MLPerf configuration files accompanying the SMC submissions, and from the experimental setup documentation for the BNL runs. Our final data set contains 116,646 individual node-power draw observation across 9 different workloads. We performed the regressions in R, with the relevant code and our raw data available in [repository link]. To evaluate model predictive performance, we estimated the power draw of workloads and configurations not represented within our training set: a later submission to ML perf for LLM training by a different submitter, and a UNET medical image classifier within the SMC submission.

*Model selection and validation*

Given the non-linear nature of our model specifications, a direct estimation approach presented convergence challenges. Attempting to simultaneously estimate all parameters across both the asymptotic and sigmoid functional forms resulted in inconsistent convergence, a common issue in non-linear regression with interdependent parameters. Our non-linear model specifications present parameter identification challenges due to strong interdependence between shape parameters ($a$, $k$, $x_0$) and magnitude parameters ($P_{idle}$, $P_{comp}$). When estimated simultaneously, these parameters often result in multiple local optima with similar fit quality but dramatically different parameter values. This "ridge" in the optimization landscape leads to unstable convergence and physically implausible parameter estimates.

To address this challenge, we implemented a two-stage estimation procedure: first constraining magnitude parameters to physically measured values while estimating shape parameters, then using those stabilized shape parameters to refine our final model. This approach prioritizes parameters with direct physical interpretation while ensuring numerical stability. Our sequential fitting strategy is analogous to profile likelihood methods, where parameters of primary interest are estimated by constructing a profile of the likelihood function over nuisance parameters [66]

We incorporated some physical constraints into our initial model based on values measured in Latif et al. (2025) for idle power and the GPU burn. The idle power for an air cool node with a rear-door heat exchanger at 24°C was measured to be a stable value of 1.8 kilowatts. During a combined stress test of the GPUs and CPUs on the node at their maximum utilization, the node stably drew 8.4 kW. For the bounded asymptotic model, we constrained the maximum power at 8.4 kilowatts, based on the empirical measurements from the GPU burn test. This represents an experimental proxy for the maximum possible utilization workload. With these constraints established, we proceeded to estimate the remaining parameters: the steepness coefficient ($a$) for the asymptotic model and the inflection point ($x_o$) and steepness ($k$) for the sigmoid model. This approach reduces dimensionality improving parameter identifiability.

To evaluate generalizability, we employed a three-part validation strategy:

1. Cluster robustness tests using heteroskedasticity-consistent standard errors grouped by workload type to account for within-workload correlation
2. Leave-One-Out Cross-Validation (LOOCV) to assess steepness parameter stability across different architectural paradigms and training configurations
3. Out-of-sample validation of final model specifications using out-of-sample workloads including an independently submitted MLPerf LLM training run and a UNET medical image classifier

To validate our estimates, we performed a cluster robustness test, treating each individual workload as a distinct cluster. This accounts for potential correlation within similar workloads while allowing independence between different architectural paradigms and configurations. The coefficient estimates for our population-level steepness parameters were cluster-robust significant (P < .01) as shown in table 1, confirming that our parameter estimates remain robust when accounting for within-workload error correlation.

To further validate the robustness of this approach, we conducted Leave-One-Out Cross-Validation (LOOCV), systematically excluding each workload from our dataset and re-estimating the steepness parameters. This procedure assesses how sensitive our parameter estimates are to the inclusion or exclusion of specific workloads. The full LOOCV summary table showing parameter stability as individual workloads are excluded is available in Appendix D.

**Table 1: LOOCV Coefficient Stability and Cluster-robust Significance**

| Cluster Robustness Results | Asymptotic Model | Sigmoid Model | |
|---:|---:|---:|---:|
| | $\alpha$ | $k$ | $x_o$ |
| Parameter Estimate | 5.55 | 1.12 | 11.46 |
| Standard Error | 1.31 | 0.09 | 0.68 |
| t-value | 4.25 | 3.6 | 16.66 |
| p-value | <2e-16 *** | 2.93*10^-7*** | <2e-16 *** |
| **LOOCV Summary** | | | |
| | $\alpha$ | $k$ | $x_o$ |
| Mean | 5.56 | 1.01 | 11.24 |
| StdDev | 0.43 | 0.4 | 0.72 |
| CoV (%) | 7.7 | 39.6 | 6.4 |

The LOOCV results demonstrate strong parameter stability across most workloads, with coefficients of variation (CoV) of 7.7% for the asymptotic model's steepness parameter and 6.4% for the sigmoid model's inflection point. However, we identified BNL - Resnet 1 as a potential outlier, with parameter estimates diverging notably from the rest of the dataset, indicating a problematically high-influence cluster within our sample.

This workload exhibited low GPU utilization: 36% on average as compared to 77-93% in the other BNL runs [46]. This is likely due to the small batch size causing frequent memory transfers that throttle GPU performance [67], [68]. This creates a fundamentally different power-computation relationship than our model aims to characterize, as the hardware is not compute-bound but memory-bound. Including this outlier would bias our model toward a power-computation relationship not representative of the performance-optimized configurations typical in production environments

Given the high coefficient stability across the other workloads, we excluded the BNL - Resnet 1 workload from our final parameter estimation to avoid biasing our model toward this atypical operating condition with a known experimental confound. Additionally, the SMC - Llama-70B (8) workload shares identical architectural, hyperparametric, and nodal characteristics with one of our validation workloads. To avoid data leakage, we excluded this workload from our training set, including it instead in the validation group. After these exclusions, we recalculated the population values for steepness parameters to use in subsequent estimation phases: $\alpha$ = 5.11 for the asymptotic model and $x_o$ = 9.91. Given the high CoV of *k*, we elected to fit our sigmoid models leaving *k* unconstrained, and they were able to successfully converge to physically reasonable values.

With stable steepness parameters identified through cross-validation, we proceeded to estimate three final model specifications. We fit dataset-wide asymptotic and sigmoid models using the established steepness parameters. Additionally, to account for architecture-specific power behaviors observed in preliminary analysis, we developed an extended asymptotic specification incorporating fixed effects for model architecture (transformer vs CNN). This architecture-specific model maintains the same functional form but allows separate magnitude parameters for each architecture type, potentially capturing the distinct power profiles which vary by the underlying model architecture [40].

# Results

## Workload Characterization

Resnet [69] models are convolutional-neural-nets (CNN), while Llama2 [38] and GPT-3 [27] are each Transformer architecture DNNs. All workloads sourced from Latif et al. are preceded by the identifier BNL, and all other workloads were sourced from SMC. The BNL workloads include two configurations of Resnet, which differ by batch size. BNL-Resnet 1 refers to the 512-image global batch workload, while BNL-Resnet 2 corresponds to the 4,096 batch size. The SMC training runs concluded upon reaching MLPerf-specified quality targets. For image classification, ResNet training continued until achieving 75.90% accuracy on ImageNet. The GPT3 language model trained until reaching a log perplexity of 2.69, while Llama-70B is fine-tuned with the LoRA method on the SCROLLS GovReport dataset concluded at a cross entropy loss of 0.925.  A full inventory of summary statistics for each workload is shown in Table 2. Boxplots of all sampled power measurements for each workload are displayed in Figure 1.

Table 2 Summary statistics for each individual workload in our dataset. All systems use NVIDIA H100-SXM5-80GB GPUs (8 per node). B = billion parameters, M = million. Batch refers to the global batch size. For each workload, average power ($P_{avg}$), maximum power ($P_{max}$), and standard deviation (SD) are reported in kilowatts (kW). Power measurements are per-node, energy values represent total IT system consumption including estimated interconnect power. Global batch size (Batch) values are sourced from the relevant configuration files. Flops are the estimated floating point operations per iteration, calculated according to equations 10-12, and Flops / node is our effective flops-per-node computational demand estimate. BNL indicates Brookhaven National Laboratory data, while SMC indicates open-source benchmarking data from Sustainable Metal Cloud. Columns proceed left-to-right from workload configuration and identifiers to measured values.

| Workload | Arch. | Param. | Batch (global) | Nodes | Flops | Flops/node | $P_{avg}$ (kW) | $P_{max}$ (kW) | SD (kW) | Duration (hours) | IT Energy (kWh) |
|---|---|---|---|---|---|---|---|---|---|---|---|
| SMC - GPT3-175B (64) | DNN | 175B | 2048 | 64 | $6.01 \times 10^{18}$ | $9.40 \times 10^{16}$ | 7.67 | 8.45 | 0.73 | 0.95 | 464.94 |
| SMC - Llama-70B (64) | DNN | 70B | 64 | 64 | $1.47 \times 10^{17}$ | $2.29 \times 10^{15}$ | 5.91 | 8.76 | 1.52 | 0.03 | 12.10 |
| SMC - Llama-70B (8) | DNN | 70B | 8 | 8 | $1.83 \times 10^{16}$ | $2.29 \times 10^{15}$ | 7.73 | 8.73 | 1.11` | 0.09 | 5.43 |
| SMC - Llama-70B (1) | DNN | 70B | 8 | 1 | $1.83 \times 10^{16}$ | $1.83 \times 10^{16}$ | 6.78 | 7.32 | 1.05 | 0.5 | 3.39 |
| BNL-Llama-13B (1) | DNN | 13B | 8 | 1 | $6.03 \times 10^{18}$ | $6.03 \times 10^{18}$ | 7.79 | 8.42 | 0.61 | 8 | 62.36 |
| SMC - ResNet 1 (8) | CNN | 26M | 3200 | 8 | $3.46 \times 10^{13}$ | $4.32 \times 10^{12}$ | 6.36 | 7.89 | 1.66 | 0.07 | 3.75 |
| SMC - ResNet 2 (1) | CNN | 26M | 3200 | 1 | $3.46 \times 10^{13}$ | $3.46 \times 10^{13}$ | 6.76 | 6.88 | 0.29 | 0.22 | 1.51 |
| BNL - Resnet 1 (1) | CNN | 60M | 512 | 1 | $3.54 \times 10^{11}$ | $3.54 \times 10^{11}$ | 4.6 | 5.02 | 0.34 | 26.8 | 123.41 |
| BNL - Resnet 2 (1) | CNN | 60M | 4096 | 1 | $2.83 \times 10^{12}$ | $2.83 \times 10^{12}$ | 5.76 | 6.48 | 0.11 | 5.25 | 30.15 |

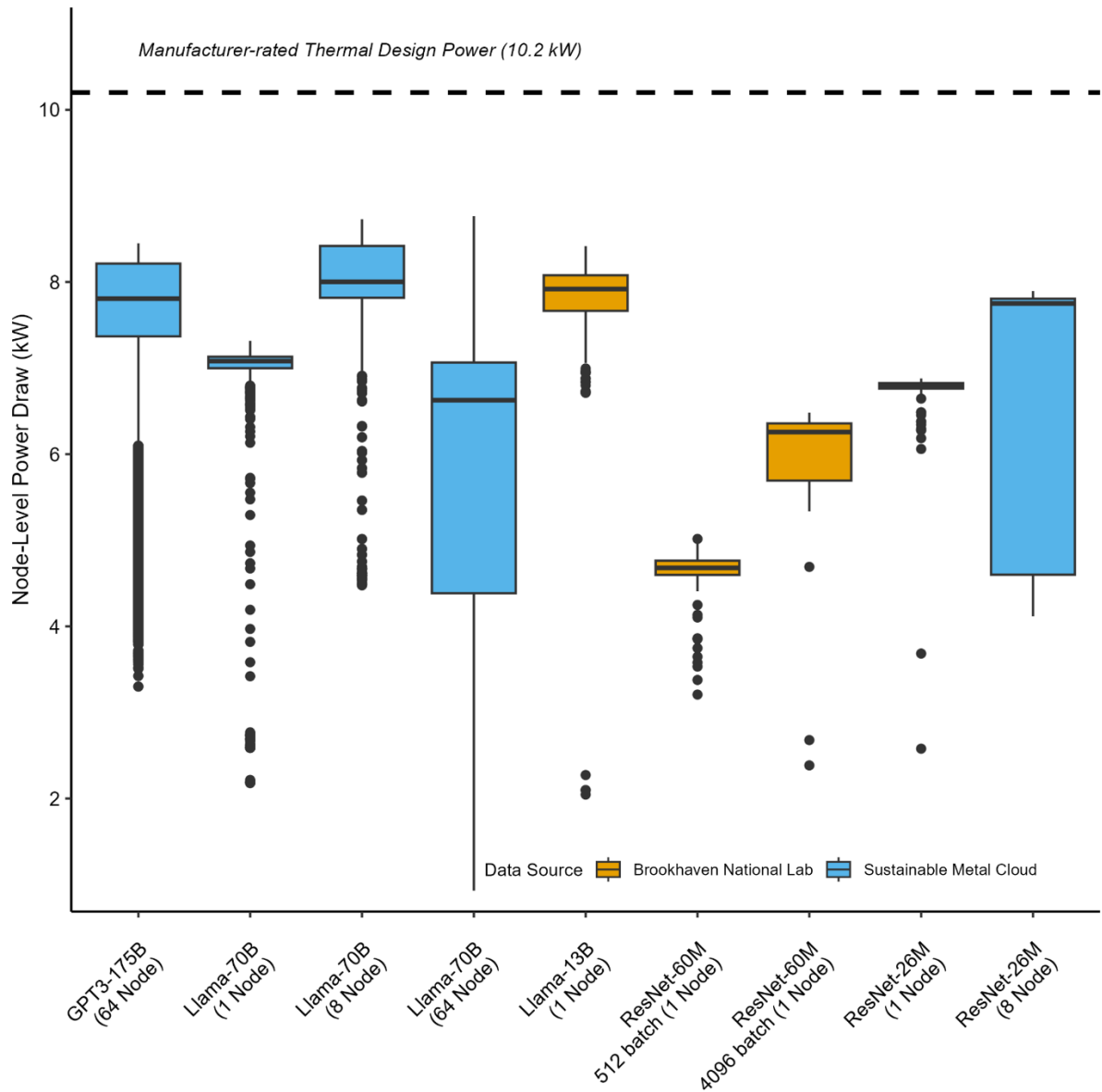

*Figure 1: Box plots of measured server power demand across different AI training workloads from Brookhaven National Lab (gold) and Sustainable Metal Cloud (blue) facilities. The dashed line indicates manufacturer-rated maximum power (TDP) of 10.2 kW. Boxes show interquartile range with median line; whiskers extend to 1.5x IQR; points indicate outliers. Note that measured power draw consistently remains well below TDP, with architectural differences evident between the deep neural net (Llama, GPT3) and shallow neural net (ResNet) workloads.*

While multi-nodal training produced the highest measured per-node maximum power, the node-level average for the most computationally dense workloads was consistent across single and multi-nodal training. In fact, the highest average node power (7.79 kW) was observed in the single node BNL training of Llama-13B. Note that the average and maximum power draw measured in each workload never approached the manufacturer rated maximum of 10.2 kW.

Models of different architectures differ in their load profile. Shallow CNNs display a flat power demand time-series, with a single, uninterrupted period of high computational intensity. This is consistent with the load profile of physical simulation and other conventional HPC workloads [70]. An illustrative example of this load shape is the image classifier training shown in Figure 2. In contrast, DNN models display characteristic square-wave load, with periods of higher demand punctuated by troughs at both the chip and the node level, as shown in Figure 3.

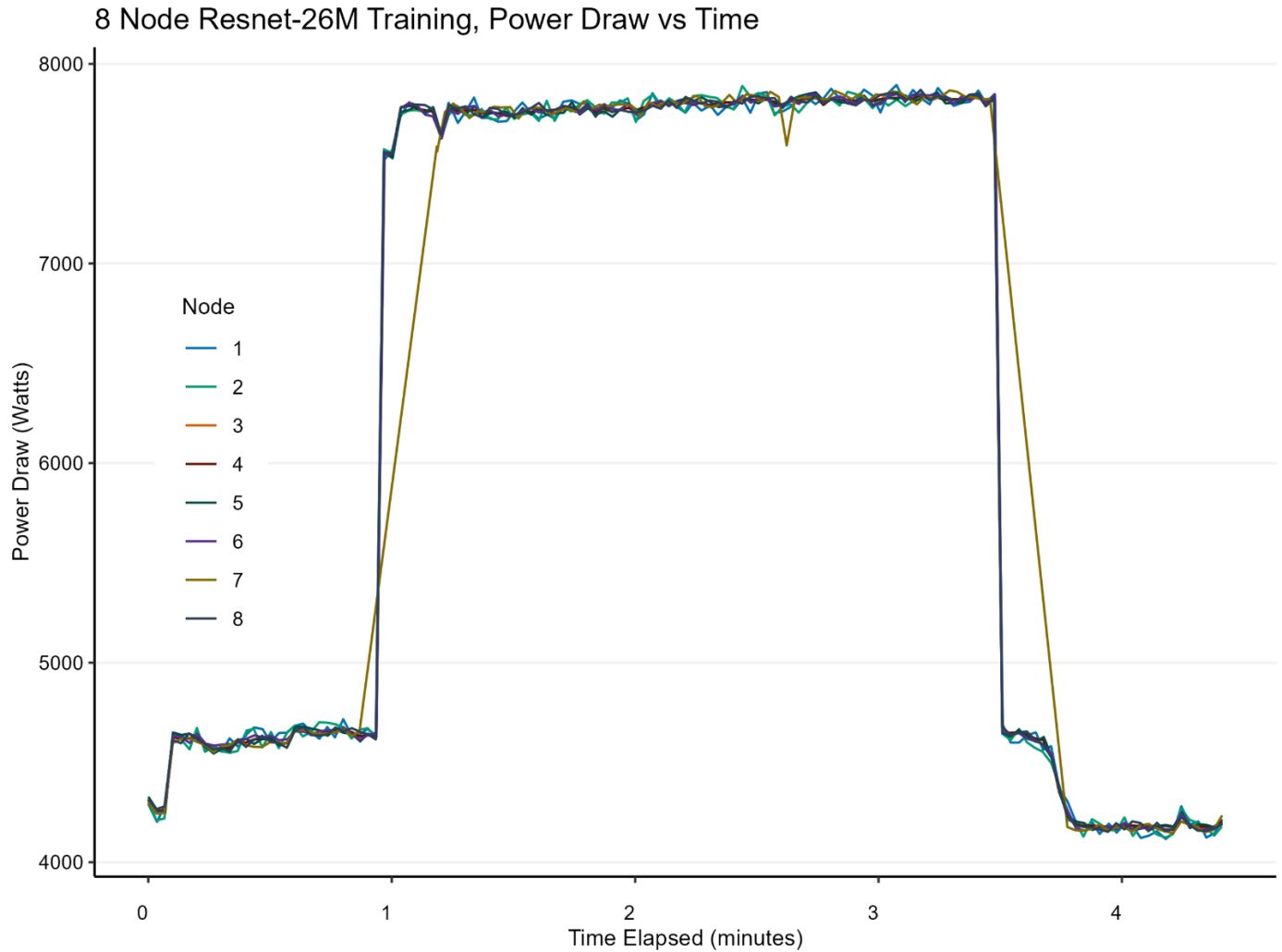

*Figure 2: Node-level power demand during 8-node ResNet-26M training. Each line represents a unique node in the training cluster. The flat power profile during the compute-intensive phase (1-4 minutes) is characteristic of CNN architectures. Note the three distinct phases: initialization (~2.5kW), compute-intensive training (~5.8kW), and completion. Minimal node-to-node variability similar to classical HPC and physical modeling workloads.*

The characteristic DNN waveform reflects the parallelization of the computation. During training, each accelerator only operates on a partition of the model using a subset of the training data. While individual GPUs can achieve high utilization during active computation, they briefly enter lower-power states while waiting for results from other parts of the system. These troughs in power demand occur at multiple scales. At the chip and node level, GPUs may be throttled while waiting for results from slower computations elsewhere in the network. System-wide drops in power draw occur during key synchronization points: when partial results are checked for consistency across nodes, when parameter updates are propagated through the system, when updated weights are written to memory, and when new training data is loaded from storage into GPU memory. Future empirical research should collect detailed component-level sub-metered demand data during training runs to identify the composition of demand over time.

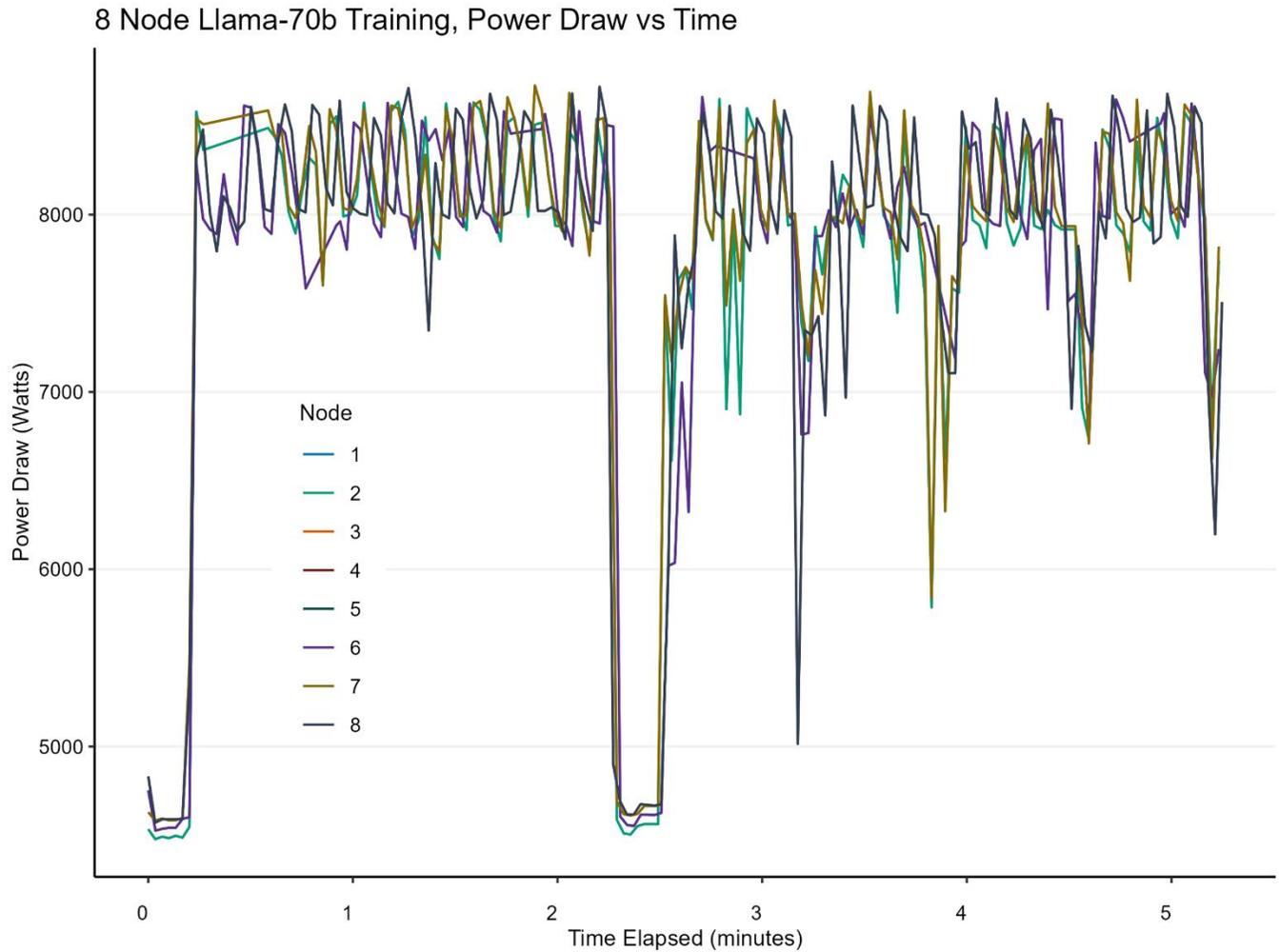

*Figure 3: Time series of node-level power demand during Llama-70B training across 8 nodes at Sustainable Metal Cloud. Each line represents a unique node. The characteristic square-wave pattern of transformer architecture training is evident, with regular drops in power demand across all nodes (e.g.at 2:20) corresponding to synchronization points and memory access periods. Power drops of individual nodes during compute intensive phases likely correspond to throttling as GPUs or nodes await intermediate values from other processors.*

These results demonstrate the importance of empirical validation of training power demand. At the node level, the interactions between accelerator computational demand, memory, supervisory logic, and interconnect mean IT power demand is not steady state within deep learning workloads. Further, previous work has found that the discrepancy between empirically measured power demand on a wattmeter and software approximation differed according to accelerator [71], [72]. The optimization of node-power-performance vs. throughput vs. system energy performance is non-trivial [73]), see Appendix A for more detail These relationships evolve as the hardware does, as clusters grow and individual racks approach 100 kW of rated IT load [51] detailed empirical understanding of these interactions should keep pace. Component level data could enable system optimization, and detailed time resolution of power draw data could enable better assessment of the grid impacts of these facilities.

## Statistical Estimators and Model Performance

Our analysis evaluated three model specifications: a basic asymptotic model, an architecture-specific model differentiating between transformer and CNN architectures, and a sigmoid model. All models were fitted using the two-stage estimation procedure described in the methods, with fixed steepness parameters derived from our preliminary analysis: α = 5.11, $x_0$ = 9.91, and a $P_{idle}$ value of 1.86 kW. Table 3 presents coefficient estimates and statistical significance for each model.

*Table 3: Robust weighted regression estimates of active power coefficients estimates, statistical significance, and prediction accuracy for three model specifications. The Asymptotic model uses a single power saturation parameter, while the Architecture-Specific model employs separate parameters for LLM and CNN workloads. The Sigmoid model implements a different functional form with transition steepness parameter k. All models were validated against four out-of-sample workloads. All significance and standard errors are the cluster-robust values. Stars denote significance codes, MAPE values show in-sample error followed by out-of-sample error in parentheses and demarcated by a dagger. Note the parameter k is dimensionless and is displayed with the others for convenience.*

| Model | COEF. | ESTIMATE (kW) | STD. ERROR (kW) | T-VALUE | P-VALUE | ENERGY MAPE (%) |
|---|---|---|---|---|---|---|
| **ASYMPTOTIC** $\widehat{Y} = P_{idle} + \beta_{comp} \dfrac{log(x)}{\alpha + log(x)}$ | $\beta_{comp}$ | 6.65 | (.35) | 19.04 | < 0.001*** | 11.8 (7.6†) |
| **ARCHITECTURE-FIXED EFFECT** $\widehat{Y} = P_{idle} + \beta_{arch} \dfrac{log(x)}{\alpha + log(x)}$ where $\beta_{arch}$ depends on model architecture | $\beta_{comp\ LLM}$ | 6.89 | (.50) | 13.65 | < 0.001*** | 11.1 (5.4†) |
| | $\beta_{comp\ CNN}$ | 6.28 | (.32) | 19.46 | < 0.001*** | |
| **SIGMOID** $Y = P_{idle} + \beta_{comp} \dfrac{e^{(x-x_0)/k}}{1 + e^{(x-x_0)/k}}$ | $\beta_{comp}$ | 6.94 | (1.33) | 5.23 | < 0.001*** | 9.3 (5.5†) |
| | k | 0.19 | (.16) | 1.19 | .235 | |

To evaluate predictive accuracy, we assessed our models using both in-sample workloads and four out-of-sample validation workloads representing diverse configurations: two 8-node Llama-70B implementations (from Dell and SMC) and UNet-3D medical image classifier workloads trained on either 1 or 9 nodes. This validation dataset spans both transformer and CNN architectures, includes single and multi-node configurations, and incorporates data from multiple sources, providing a robust test of model generalizability. For each workload, we compared estimated energy consumption (based on modeled power and average node duration) to empirically measured energy use to calculate Mean Absolute Percentage Error (MAPE). A full table of summary statistics for the validation workloads is available in Appendix D.

While the sigmoid model achieved the best in-sample fit (MAPE = 9.34%), the architecture-specific asymptotic model demonstrated superior generalization to unseen workloads (MAPE = 5.39%). This suggests that accounting for architectural differences in power consumption patterns can improve workload power prediction. The superior out-of-sample performance of both the sigmoid and architecture-specific models compared to in-sample accuracy likely reflects the validation datasets' closer alignment to compute-optimal configurations. Overall, coefficients were stable across different specifications, with estimates differing by less than cluster robust standard errors. Note the comparatively large standard error in sigmoid active power estimate, and the lack of cluster-robust significance in the steepness estimate.

To assess model robustness, we conducted Leave-One-Out Cross-Validation (LOOCV), systematically excluding each workload and re-estimating model parameters. Table 4 summarizes the stability metrics across all models.

Table 4: Parameter stability analysis across model specifications, showing means, coefficients of variation (CoV), and ranges when systematically excluding each workload. Lower CoV values indicate greater parameter stability.

| MODEL | PARAMETER | MEAN (KW) | COV (%) | RANGE (KW) |
|---|---|---|---|---|
| **ASYMPTOTIC** | $\beta_{Comp}$ | 6.6 | 2.0 | 6.4-6.8 |
| **ARCHITECTURE-SPECIFIC** | $\beta_{Comp\ LLM}$ | 6.9 | 3.2 | 6.6-7.3 |
|  | $\beta_{Comp\ CNN}$ | 6.3 | 2.5 | 6.0-6.6 |
| **SIGMOID** | $\beta_{Comp}$ | 6.7 | 8.6 | 5.4-7.6 |
|  | $k$ | 0.23 | 42.1 | 0.14-0.46 |

All asymptotic model variants demonstrated excellent parameter stability with coefficients of variation (CoV) below 3.2%, indicating robust estimation regardless of which workload was excluded. The architecture-specific parameters maintained tight ranges despite architectural diversity in our dataset, with LLM compute power (β_compute_llm) consistently higher than CNN power (β_compute_cnn) across all LOOCV iterations. The sigmoid model showed acceptable stability for P_max (CoV = 8.59%) but poor stability for the transition steepness parameter k (CoV = 42.14%), confirming parameter identification challenges in this specification.

Based on our analysis of cluster-robustness, predictive validity, and parameter stability, we selected the architecture-specific asymptotic model as our preferred specification. This model balances excellent parameter stability (CoV < 3.2%) with superior out-of-sample prediction accuracy (5.39% MAPE), while capturing power consumption differences between transformer and CNN architectures. The basic asymptotic model offers slightly better parameter stability but sacrifices prediction accuracy and fails to capture architectural differences in power behavior. The sigmoid model, despite strong predictive performance, suffers from parameter instability and identification issues, and a lack of cluster-robust significance in the steepness parameter, that limit its practical utility.

In terms of the performance of the statistical estimator, our goal was to better predict the total energy use of these workloads using this estimate of average node power demand while training. To evaluate this, we compare our empirically measured energy use for each workload to an estimate based on training duration and an approximation of node power. These approximations include our architecture fixed-effect model prediction, the TDP of the GPU as a lower bound (GPU-Hours X GPU TDP), and the TDP of the Node as an upper bound. The percentage errors of these three estimates of node power draw across our in-sample data are as follows: $\delta_{node}$= 36.83%, $\delta_{chip}$= 27.26%, $\delta_{model}$= 11.46%. In terms of practical utility, estimating workload energy use provides the most value. We evaluated each model's predictive performance using Mean Absolute Percentage Error (MAPE) for both in-sample and out-of-sample predictions. Plots of the architecture-fixed-effect estimates of total energy use of both in-sample and out-of-sample workloads, normalized to the empirical measurements for each workload are shown below in Figure 5.

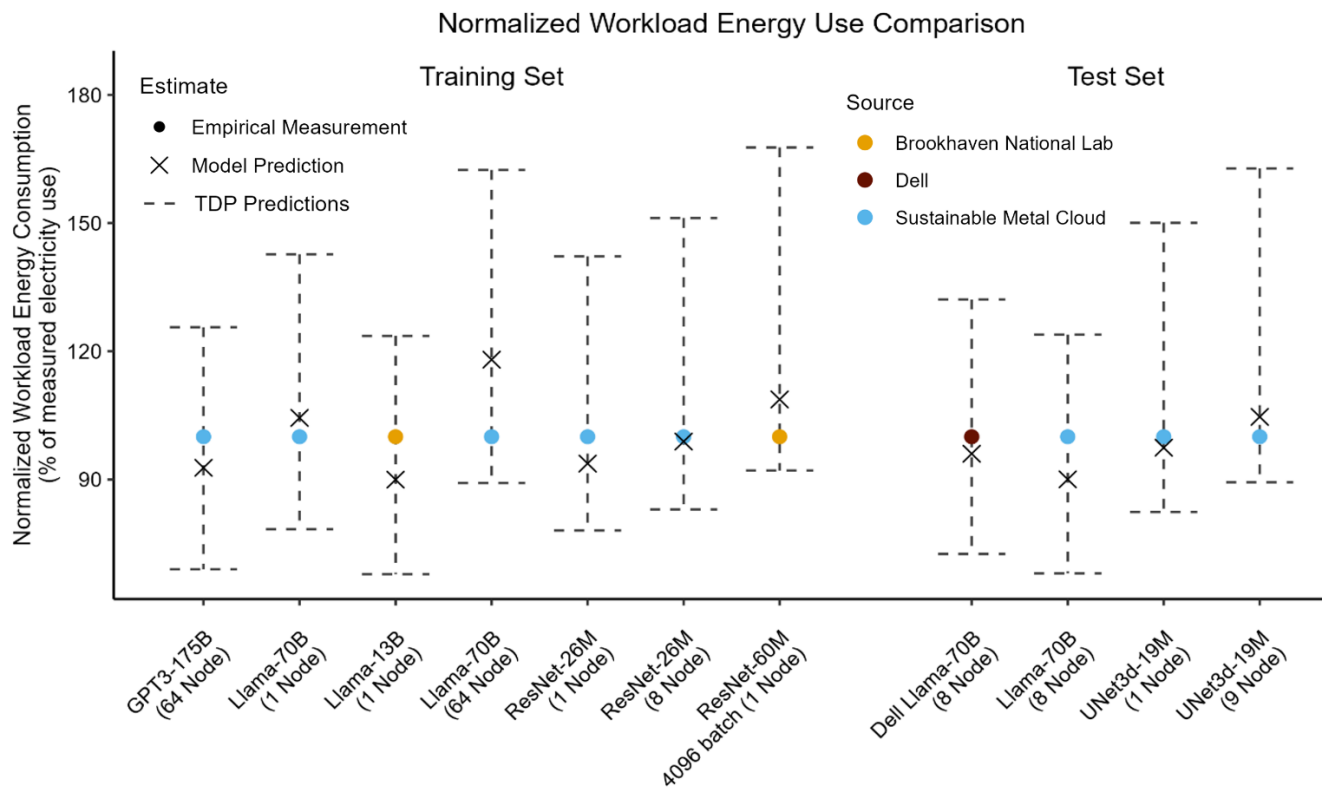

*Figure 4: Comparison of workload energy use estimates normalized to empirical measurements (100%). Dotted bars span the range between chip TDP-based (lower bound) and node TDP-based (upper bound) estimates. Solid dots indicate empirically measured energy consumption, color according to the data source, black X markers indicate model-predicted values. In-training sample workloads are on the left of the plot, while out-of-sample workloads are on the right of the plot. The proposed architecture-fixed-effect model energy estimates (mean absolute percent error 11.05% in sample, 5.39% out of sample) outperforms both TDP-based approaches (chip TDP MAPE 21.82%, node MAPE error 42.41%) across evaluated workloads. Out-of-sample performance likely reflects workloads with higher computational saturation.*

## Limitations

A main limitation to this research is data breadth and granularity. The largest commercial models are orders of magnitude higher parameter count and training by tens of thousands of individual GPUs. While our workloads appeared to saturate available compute resources, larger-scale workloads and training clusters may exhibit distinct behavior. Hyperparameter choice, exact hardware configuration, and cooling system can each affect training power behavior, and future research should extend data collection efforts to consider these characteristics. Prior work has identified systematic discrepancies between GPU-embedded sensor readings and externally validated power measurements using calibrated instrumentation [71], [74]. While facility-level power monitoring through DCIM software and PDU metering typically provides more reliable data than onboard sensors [65], the temporal resolution and component-level granularity of these measurements could be improved through more sophisticated metering.

The interaction between IT load and cooling infrastructure represents another limitation. We prioritized estimating IT load as in energy-systems applications, infrastructure and facility energy-use are typically extrapolated from this parameter. However, IT load and infrastructure are not fully independent, especially in compute-dense applications [75]. H100 GPUs operate with dynamic-voltage-frequency-scaling (DVFS), where voltage and clockspeed are adjusted to optimize energy or throughput based on workload characteristics and chip temperature [76]. This creates a bidirectional relationship between cooling efficiency and computational performance: more effective cooling enables higher operating frequencies and more energy-efficient transistor operation [77]. DVFS techniques can dramatically improve energy performance [78], with inference representing a particularly promising application [79], [80].

Direct-to-chip liquid cooling or immersion cooling could enable different power-performance tradeoffs by maintaining lower junction temperatures. This would allow GPUs to sustain higher clock frequencies without thermal throttling, potentially resulting in higher instantaneous power draw but shorter training times and potentially improved overall energy efficiency. Future work should systematically evaluate the impact of cooling technology on node-level power profiles across identical workloads, ideally using controlled experiments with the same computational hardware under different cooling regimes. Additionally, the operational intensity of a given model (required FLOPS per parameter) is known to vary according to architectural choices. More sophisticated model specifications based on detailed relationships between hyperparameter choice, architectural characteristics, and workload configuration may enable more accurate prediction, conditional on their ability to generalize across workloads. The lack of component-level submetered power data also limits our ability to precisely attribute power draw to specific elements of the system during different phases of training. These limitations should be considered when assessing the implications of this work.

## Discussion

To illustrate the importance of accurate power measurement, consider a hypothetical training cluster in Iowa containing 20,000 H100 GPUs across 2,500 nodes—a scale consistent with current commercial clusters [81]. Consider all 2,500 nodes to be training foundation model for 90 days, rumored to be the approximate training duration of current commercial models. Assume the cluster has a PUE of 1.1, the value Alphabet reported for their facilities in 2023 [82], and ten percent facility power conversion and switching losses.

Our architecture fixed-effect asymptotic model predicts each node draws 7.3 kW. Over the 90-day training period, we estimate that this cluster would consume approximately 49.0 GWh of electricity. Using Iowa's regional grid carbon intensity of 428 kg CO2e per MWh [83], this training run would then emit 20,970 tons of CO2e. However, if we had instead relied on the manufacturer-rated TDP of 10.2 kW per node, as many facility planners do, our estimate of total energy consumed would be 68 GWh. This 19 GWh difference in estimates represents approximately 8,132 tons of CO2e - equivalent to the annual emissions of 1,900 passenger vehicles.

The carbon intensity value itself introduces considerable uncertainty. Using the EPA's non-baseload value of 806 kg CO2e per MWh rather than the average produces estimates differing by 15,300 tons CO2e. Other accounting methods—estimated marginal emissions, time-based hourly emissions, or system expansion emissions—would yield different results entirely. This demonstrates why accurate power measurement proves critical for both infrastructure planning and environmental impact assessment, as measurement uncertainty compounds across other uncertain impact estimates. While this hypothetical facility's absolute energy and carbon footprint remains significant regardless of the calculation method, effective planning and optimization demands precise operational understanding.

Uncertainty in the IT load for AI training facilities is not merely a theoretical concern; it creates tangible challenges for infrastructure providers responsible for electrical grid stability and capacity planning. The lack of transparency around actual power demand in the data center sector has historically resulted in numerous inaccurate claims and projections of energy consumption [84].  Empirical data to understand IT power demand under actual operating conditions was a key need highlighted in a recent report requested by U.S. Congress to strategically meet the demands of future data center energy use [85]. In our conversations with electrical utilities and energy infrastructure planners, uncertainty in actual IT demand consistently emerged as the primary pain point when working with these facilities. As one grid infrastructure planner candidly expressed: "The most significant headache in working with [AI training facilities] is that I don't know how much [electricity] their IT will actually consume...when a facility claims it's 100 MW [rated IT load], how much demand is that really?"[4] This ambiguity compounds through multiple sources of uncertainty: the gap between rated and actual hardware power consumption, facility-specific cooling configurations (which can vary from traditional air cooling to more efficient direct-to-chip liquid solutions), and phased deployment schedules that may spread capital expenditures across years. Infrastructure planners must determine whether a facility marketed as "100 MW of IT load" represents fully deployed capacity or includes planned expansions financed by initial operations. These uncertainties directly impact critical planning decisions: transformer and substation capacity requirements, transmission infrastructure sizing, and even generation resource adequacy assessments.

By providing empirically-validated models of current-generation AI hardware power consumption during training workloads, our research helps reduce this uncertainty, enabling infrastructure providers to calibrate their planning assumptions against real-world measurements. This prevents both costly overprovisioning of capacity and the risks associated with inadequate supply, which could result in service interruptions or grid stability issues. The rapid expansion of AI infrastructure need not jeopardize energy system reliability or the clean energy transition, but it requires both greater transparency in facility planning and careful management of actual operational demand profiles. The sheer scale of the workload means that operators are well incentivized towards efficiency [86]. Additionally, the up-front capital requirement and small supply of AI training hardware have resulted in a concentration of capabilities in the largest scale data center operators [84], service providers with demonstrated industry leading efficiency performance [87], [88], [89]. The heat density of specialized AI training chips makes them ideal matches for the highest efficiency, cutting-edge liquid cooling systems [90], both direct-to-chip and immersion. Recognizing this reality, in 2023 Meta fundamentally redesigned their AI data center for liquid cooling infrastructure [91]. The duration of large-scale training runs means that AI training has low latency sensitivity [29]. This is not to say that individual elements of the AI training hardware stack are latency tolerant between one another, rather that the duration of training runs makes them insensitive to location derived network delays. This robustness combined with the post-training processing models go through means that these facilities have high degrees of freedom in site selection. As the bulk of data center environmental impact is site dependent [21], AI training centers are free to locate where the power is green and inexpensive, can be carbon-aware scheduled [29], and are strong candidates for demand response [92].

The relationship between cooling technology and energy efficiency requires some detailed consideration. While liquid cooling enables higher GPU operating frequencies and voltages, this can result in non-linear increases in power consumption due to the voltage-frequency characteristics of semiconductor devices. At the node level, liquid cooling may reduce energy consumption by replacing inefficient air-cooling fans, but the net effect on system energy use is complex. Users may leverage the increased computational throughput enabled by liquid cooling to run more intensive workloads, potentially creating a rebound effect in energy consumption. Alternatively, while liquid cooling might enable operating states with worse instantaneous energy efficiency (in terms of operations per watt), the associated performance improvements could reduce total training time and improve lifecycle energy use. Understanding these tradeoffs requires detailed empirical measurement, at the chip, node, cluster, and facility level.

---

[4] Personal communication, Virginia electrical utility employee, April 2025

Previous work projecting large growth in data center energy use did not adequately account for the improvements data center energy efficiency, which enabled global internet traffic to grow by 550% from 2018, while data center energy use only went up 6% [88]. However, AI compute is fundamentally different from traditional high-performance computing, and counting on data center efficiency to blunt the growth of AI energy use is not a robust strategy. The stakes for accurate forecasting are high - future data center electricity projections inform stakeholders on needed development for electricity generation and transmission to allow for strategic growth that benefits the environment and broader economy [93]. Overestimating future electricity demand may lead to development of electricity infrastructure unable to recoup costs, while underestimates can impede AI technology progress. AI electricity growth scenarios by LBNL (Lawrence Berkeley National Laboratory) indicate that understanding AI power demand during operation relative to hardware rated power is a key uncertainty that greatly affects our understanding of future AI electricity needs [85].

Our results suggest that servers for AI workloads are generally not demanding power up to their manufacturer rated power during model training. This likely results in AI energy use to be slightly overestimated, at least for the time being. Yet, the seemingly inelastic demand for AI compute will likely erode gains from computational efficiency. In addition, our findings that AI training workloads fluctuate rapidly in power-demand at the node level, results consistent with findings at the chip level [31], would introduce new stresses to the grid in requiring fast-ramping and grid-forming resources to ensure grid stability [94]. These workload timeseries suggest that an 80,000 chip cluster would have in-training demand swings of approximately 24 MW in the space of 20 ms. Swings of such scale and speed are known to cause issues for local electric grids at HPC centers [95]. Both the potential growth in data center energy use and the variability in power demanded provide a rationale for high-resolution data transparency from firms of their AI data center energy use. These data would enable data center users to optimize and smooth power demands and grid operators to manage anticipated loads.

In contrast to prior work from the machine learning community which largely focuses on the per-device GPU power measurements as a proxy for total power draw, we demonstrate that components such as: CPU, memory, and interconnect fabric, substantially contribute to the node-level power demand for training workloads. As current state-of-the-art models scale in size and necessitate distribution across multiple nodes with greater reliance on internode memory fabrics, it is necessary for ML researchers and practitioners to characterize the power draw of this infrastructure for a holistic understanding of a workload's energy demands.

While reporting total energy use to train a given model is vital for assessing efficiency and hardware utilization, it is insufficient to characterize the demands these workloads place on electrical supply infrastructure. The temporal power demand characteristics, particularly high frequency demand fluctuation observed in transformer architecture training, can create significant operational challenges [96]. As models grow in parameter count and training duration, and as training potentially becomes distributed across facilities, understanding these load profiles becomes critical for infrastructure planning and operation. High-resolution time series power measurements would enable several key optimizations: identification of synchronization bottlenecks, evaluation of different parallelization strategies, and opportunities for power-aware scheduling [97]. Additionally, detailed power data could inform the co-design of algorithms and hardware to better manage these distinctive load patterns, potentially enabling more efficient facility operation without sacrificing training or model performance.

## Sources of Uncertainty in AI Energy Use Modelling

These results suggest several lines of research to develop our understand of AI related energy use. One is continued empirical effort to comprehensively characterize the energy performance of hardware running AI workloads. Benchmarking the energy performance of training across different cooling technologies is one such approach. Alternatively, detailed component-level submetering would provide more granular understanding of subsystem utilization and could potentially enable workload optimization. As hardware and facilities evolve, empirical work must keep pace. Future analyses or benchmarks of training energy should include node or cluster level power demand time series and GPU utilization [84].  While more technically challenging [98], component sub-metering would provide greater still greater insight

Machine learning research should expand to more systematically characterize the energy implications of architectural choices. The energy performance impacts of hyperparameter choice, multi-facility training, and parallelization across the 10,000 GPU clusters remain unclear. Understanding these relationships could inform model development - complementing the traditional metrics of accuracy and training time with energy performance. Energy systems research should develop more granular understanding of these facilities' load profiles and grid impacts, grounded in empirical power draw data. More detailed characterization of load variability and ramp rates would enable both better infrastructure planning and potential optimization opportunities. Every time a benchmark is run, energy and power should also be measured.

The nature and scale of demand for inference is a major source of uncertainty in long-term models of AI related energy use, with a more detailed discussion available in Appendix B. There are significant economic factors which underly scholarly projections of future inference energy use. One is the rate of consumer adoption of AI related products beginning in 2023; by some measures ChatGPT had the fastest rate of consumer adoption in software history [99]. Additionally, the training cost structure encourages expansive deployment of AI models. Recall that training is performed in discrete generational batches, typically on a hardware stack dominated by upfront capital costs [100]. This training cost structure means that firms have a strong incentive to deploy trained models as widely as possible, to better amortize the upfront training cost. Inference costs will be passed through to the user, either directly through subscription fees or indirectly through monetization of inference related web traffic. Thus, while firms will seek to recoup model training costs through broad deployment and long-term use, inference queries will be driven by end user value perception. These structural factors give us directional insight into inference computational share, but scale remains an open question, as it will be driven by user demand.

Models of future inference energy consumption face interacting uncertainties: hardware performance, hardware composition, algorithmic efficiency, demand profile, and facility characteristics. The effective energy performance of the inferencing hardware stack is in part determined by composition, what chips are doing the inferencing. The representative facility and cooling system for inference will depend on this hardware stack, as well as the latency sensitivity of the workload. Demand for inference will inform both the total scale of the workload, and the latency sensitivity of this demand. Accurately predicting any of these characteristics in isolation is daunting, assessing all four as they mutually interact presents a profound challenge. Future work on modeling the energy footprint of inference should explicitly state the assumptions of these characteristics for each scenario, to ensure a clear understanding by researchers of the respective sensitivity.

# Conclusion

Our empirical analysis demonstrates that AI training hardware operates well below manufacturer-rated power limits, with even computationally intensive workloads drawing no more than 76% of rated TDP. Through development of a novel statistical model relating computational intensity to node-level power demand, we achieved significantly improved accuracy in predicting training energy consumption compared to conventional TDP-based approaches. This more precise understanding of actual power dynamics enables better infrastructure planning and optimization while highlighting important differences between CNN and transformer architecture power signatures that must be considered in facility design.

The rapid evolution of AI capabilities and corresponding growth in computational demands makes accurate power measurement and modeling –and managing uncertainty— increasingly critical. Our findings that AI training workloads exhibit substantial power fluctuations, particularly in transformer architectures, suggests these facilities may place novel demands on electrical infrastructure. While the industry's economic incentives for efficiency and flexibility in facility siting provide some environmental advantages, the seemingly inelastic demand for AI compute could quickly erode efficiency gains. This underscores the importance of high-resolution empirical data collection and transparency from operators to enable both facility optimization and effective grid integration.

These results point to several key priorities for future research: detailed component-level power measurement to better understand subsystem interactions, systematic investigation of energy implications across model architectures and scales, and development of more sophisticated inference energy models that can account for heterogeneous hardware configurations and varying latency requirements. As AI compute demands continue to grow, such empirically-grounded understanding will be essential for managing energy system impacts while ensuring reliable infrastructure operation.

# Resource Availability

The dataset used in this research is available on the Kilthub Repository at: DOI 10.1184/R1/29067572

Additionally, a Zotero library of all sources referenced in this work can be found at: https://www.zotero.org/groups/5807620/ai_training_node-level_power_draw

# Acknowledgements

This manuscript has been authored by an employee of the Lawrence Berkeley National Laboratory under Contract No. DE-AC02-05CH11231 with the U.S. Department of Energy for the Industrial Efficiency and Decarbonization Office (IEDO). The U.S. Government retains, and the publisher, by accepting the article for publication, acknowledges, that the U.S. Government retains a non-exclusive, paid-up, irrevocable, world-wide license to publish or reproduce the published form of this manuscript, or allow others to do so, for U.S. Government purposes. Thank you to Sarah Smith, Alex Hubbard, Nichole Hanus, Dale Sartor, and Sylvia Downing.

# Author Contributions

A.C.N.: Conceptualization, Methodology, Formal Analysis, Data Curation, Visualization, Writing - Original Draft, Writing - Review & Editing

J.F.: Conceptualization, Methodology, Writing - Original Draft, Writing - Review & Editing

J.K.: Conceptualization, Writing - Original Draft, Writing - Review & Editing

I.L.: Investigation, Data Collection

E.S.: Conceptualization, Resources, Writing - Review & Editing

A.S.: Funding Acquisition, Writing - Review & Editing

C.S.: Conceptualization, Supervision, Writing - Original Draft, Writing - Review & Editing

# Declaration of interests

The authors declare no competing interests.

# Declaration of generative AI and AI-assisted technologies in the writing process

During the preparation of this work, the author(s) used *Claude* to improve language, clarity, and typesetting. After using this tool or service, the author(s) reviewed and edited the content as needed and take(s) full responsibility for the content of the publication

[87] N. Jones, "How to Stop Data Centres From Gobbling Up the World's Electricity," *Nature*, vol. 561, no. 7722, pp. 163–166, Sep. 2018, doi: 10.1038/d41586-018-06610-y.

[88] E. Masanet, A. Shehabi, N. Lei, S. Smith, and J. Koomey, "Recalibrating global data center energy-use estimates," *Science*, vol. 367, no. 6481, pp. 984–986, Feb. 2020, doi: 10.1126/science.aba3758.

[89] A. Shehabi, S. J. Smith, E. Masanet, and J. G. Koomey, "Data center growth in the United States: decoupling the demand for services from electricity use," *Environ. Res. Lett.*, vol. 13, no. 12, p. 124030, Dec. 2018, doi: 10.1088/1748-9326/aaec9c.

[90] Y. Zhang *et al.*, "Cooling technologies for data centres and telecommunication base stations – A comprehensive review," *Journal of Cleaner Production*, vol. 334, p. 130280, Feb. 2022, doi: 10.1016/j.jclepro.2021.130280.

[91] S. Moss, "Meta details AI data center redesign that led to facilities being scrapped." Accessed: Jul. 14, 2023. [Online]. Available: https://www.datacenterdynamics.com/en/analysis/meta-details-ai-data-center-redesign-that-led-to-facilities-being-scrapped/

[92] A. A. Chien, L. Lin, H. Nguyen, V. Rao, T. Sharma, and R. Wijayawardana, "Reducing the Carbon Impact of Generative AI Inference (today and in 2035)," in *Proceedings of the 2nd Workshop on Sustainable Computer Systems*, in HotCarbon '23. New York, NY, USA: Association for Computing Machinery, Aug. 2023, pp. 1–7. doi: 10.1145/3604930.3605705.

[93] Executive Order 14110, *Safe, Secure, and Trustworthy Development and Use of Artificial Intelligence,* vol. 75,195. 2023.

[94] Y. Li, M. Mughees, Y. Chen, and Y. R. Li, "The Unseen AI Disruptions for Power Grids: LLM-Induced Transients," Sep. 09, 2024, *arXiv*: arXiv:2409.11416. doi: 10.48550/arXiv.2409.11416.

[95] N. Bates *et al.*, "Electrical Grid and Supercomputing Centers: An Investigative Analysis of Emerging Opportunities and Challenges," *Informatik Spektrum*, vol. 38, no. 2, pp. 111–127, Apr. 2015, doi: 10.1007/s00287-014-0850-0.

[96] I. R. Abdulveleev, T. R. Khramshin, G. P. Kornilov, and R. R. Abdulveleeva, "Experimental Study of the Impact of a DC Electric Arc Furnace on a Power Grid," in *2021 International Conference on Industrial Engineering, Applications and Manufacturing (ICIEAM)*, May 2021, pp. 214–218. doi: 10.1109/ICIEAM51226.2021.9446320.

[97] J. Yan *et al.*, "Energy-aware systems for real-time job scheduling in cloud data centers: A deep reinforcement learning approach," *Computers and Electrical Engineering*, vol. 99, p. 107688, Apr. 2022, doi: 10.1016/j.compeleceng.2022.107688.

[98] M. Fahad, A. Shahid, R. R. Manumachu, and A. Lastovetsky, "A Comparative Study of Methods for Measurement of Energy of Computing," *Energies*, vol. 12, no. 11, p. 2204, Jun. 2019, doi: 10.3390/en12112204.

[99] K. Hu and K. Hu, "ChatGPT sets record for fastest-growing user base - analyst note," *Reuters*, Feb. 02, 2023. Accessed: Jun. 06, 2024. [Online]. Available: https://www.reuters.com/technology/chatgpt-sets-record-fastest-growing-user-base-analyst-note-2023-02-01/

[100] D. Patel and D. Nishball, "Nvidia Blackwell Perf TCO Analysis - B100 vs B200 vs GB200NVL72." Accessed: Jun. 06, 2024. [Online]. Available: https://www.semianalysis.com/p/nvidia-blackwell-perf-tco-analysis

[101] C. Miller, *Chip War: The Fight for the World's Most Critical Technology*. Scribner, 2022.

[102] T. Ben-Nun and T. Hoefler, "Demystifying Parallel and Distributed Deep Learning: An In-depth Concurrency Analysis," *ACM Comput. Surv.*, vol. 52, no. 4, p. 65:1-65:43, Aug. 2019, doi: 10.1145/3320060.
31

# Appendices

## Appendix A: AI Training: IT Hardware and Facility Overview

AI applications possess unique attributes that place novel demands on hardware infrastructure compared to previous enterprise computing workloads. State-of-the-art deep learning systems are highly parallelized, benefit tremendously from scale, and require specialized IT hardware. AI chips include graphical processing units (GPUs), field programmable gate arrays (FPGAs), tensor processing units (TPUs), and some application-specific-integrated-circuits (ASICs) specifically designed for AI workloads. In computer hardware there is a typical trade-off between generality and efficiency/speed; AI applications accentuate these tradeoffs due to the quantity of required operations. AI accelerators perform multiple computations in parallel, and conduct calculations at reduced precision without sacrificing output accuracy [8].

The most general-purpose computer hardware is the central processing unit (CPU), the workhorse of conventional enterprise computer infrastructure. While theoretically capable of the repeated predictable computations required to train and inference AI models, CPUs are at a profound performance disadvantage. Compared to CPUs, AI accelerators are ~10x-1000x faster and more energy efficient in both training and inferencing applications [8, p. 23]. These performance gains are so acute that training cutting-edge models relies on the most advanced hardware. Improvements in algorithmic efficiency, computational hardware, and system optimization have enabled higher parameter count and inference accuracy rather than energy performance [44]. Early configurations of AI training hardware were as humble as a single accelerator, by late 2024 the largest clusters contained tens of thousands of individual GPUs [81].

The most common type of AI processor is the GPU, which were originally developed to process computer graphics[5]. In the early 2010s, machine learning researchers found that they could train models faster on GPU integrated hardware [101], and by 2017 GPUs were the hardware architecture in the majority of machine learning applications [102]. GPUs are predominantly used for AI training; they are deployed to a lesser extent for inferencing. While capable of performing robust parallel computation, GPUs are general purpose devices suitable for a variety of applications. TPUs are a related type of processor designed by Google for their internal AI and ML applications. Though technically an ASIC, TPUs are closest in architecture and performance to a machine-learning optimized GPU. Like GPUs, TPUs are best suited to training workloads, though the hardware is sometimes shifted to inferencing when it would otherwise be idle or during peak inference demand [103]. The most popular and highest performance GPUs for AI training are designed by Nvidia, manufactured by Taiwan Semiconductor Manufacturing Company (TSMC), such as the H100 or B200. For the purposes of our energy use model, we selected Nvidia H100 GPUs as the representative AI training chip, owing to their dominant market share at time of writing. Note the extreme up-front capital cost to building out dedicated AI training infrastructure with these devices, which cost on the of ~$40,000 per GPU the year of their release [104].

To optimize model performance, training will iteratively pass data backwards and forwards through neural layers, adjusting weighting according to an optimization function. Additionally, individual calculations will often require input from long-range dependencies or another output, and will thus require interfacing with other neurons, layers, or values stored in memory. In practice this means that during training, an individual parameter, batch, chip, or even rack can be throttled by memory, networking, or just a slower computation elsewhere in the network. Training optimization then batches data and computations to minimize this throttling.

We can build out a description of AI training facilities based on these characteristics. The computation will take place on specialized, accelerated processors optimized for parallel operations and back propagation. As any individual computation often requires input data from another computation or memory, these processors rely on large amounts of memory and high-bandwidth interconnect [105]. To avoid significant throttling, these processors should ideally be in close physical proximity. The continued performance improvement with increased computational load [26], [27], [106] means these will be some of the largest compute scale facilities in the industry. The energy performance of a given training workload is highly sensitive to choice of accelerator. Wang et al. empirically measured the energy use to train and fine tune the AI model BERT across two different 4-GPU nodes: one with A100s, and one with RTX 8000s [71]. This is a notable comparison as both nodes have the same manufacturer rated TDP of 1,500W, but while the A100 is optimized for AI workloads, the RTX is a more general purpose chip appropriate for a variety of parallel applications such as graphical rendering. Architectural optimization in the A100 resulted in dramatic efficiency gains: training BERT on the A100 Node consumed 146 kWh, while the RTX based node required 368 kWh.

---

[5] Note that graphical processing, physical simulations, and machine learning are computationally similar. This is not coincidence; graphical processing can be conceptualized as a highly simplified physical model of materials, fluids, and optical behavior.

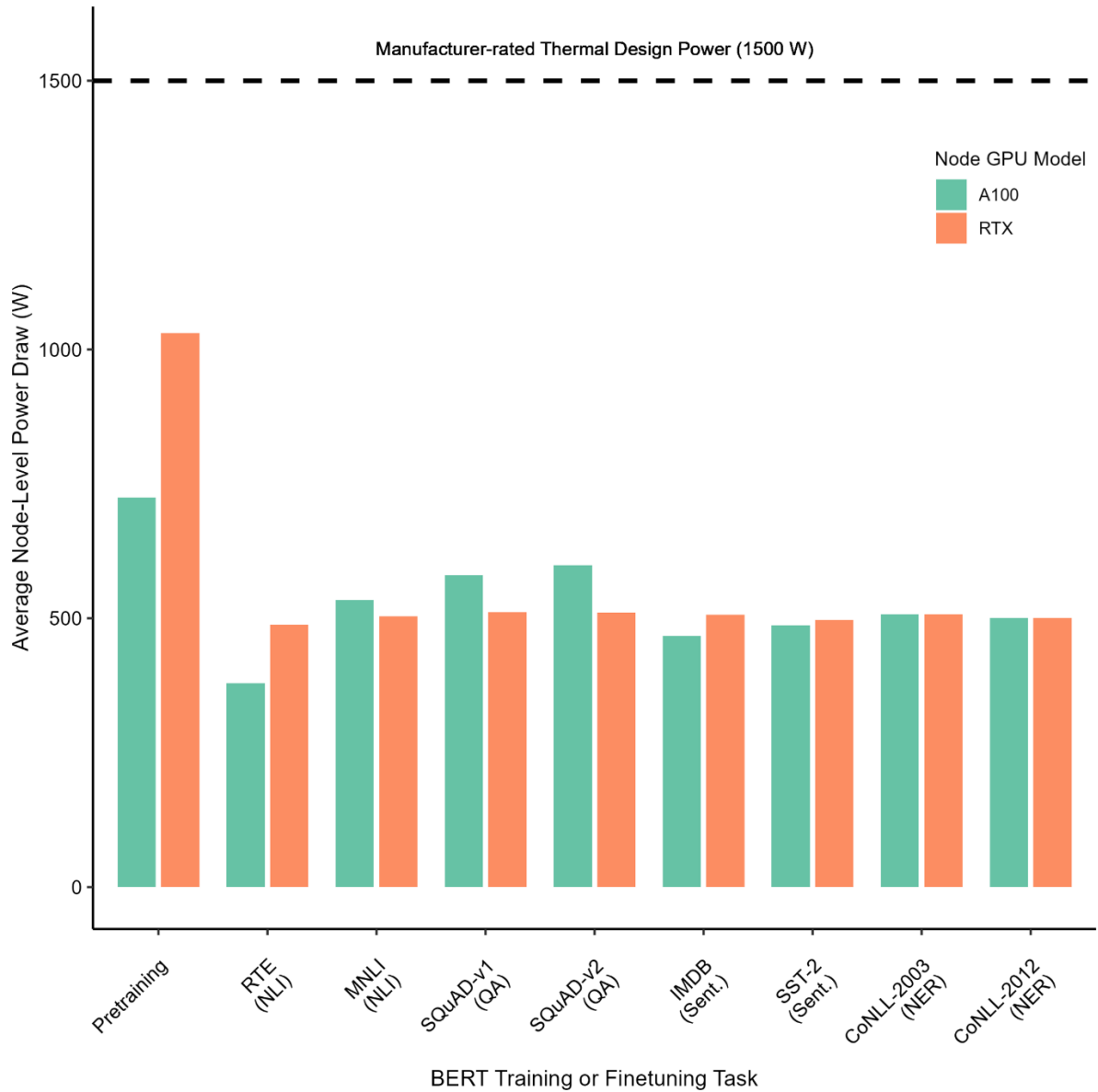

*Figure A1: Average node-level power draw during BERT model training and fine-tuning workloads across different 4-GPU node configurations (NVIDIA A100 vs RTX 8000). The dashed line indicates manufacturer-rated node TDP (1500W). Fine-tuning tasks*

*include natural language inference (NLI), question answering (QA), sentiment analysis (Sent), and named entity recognition (NER). Despite identical TDP ratings (dashed line, 1500W), the AI-optimized A100 achieves higher computational throughput at lower average power draw for the most computationally intensive workload (pretraining). The efficiency advantage of specialized AI hardware is particularly evident in the pretraining phase. Note that the duration of A100 pretraining was 202 minutes as compared to 357 for the RTX based node, resulting in significantly reduced pretraining energy use. Data sourced from Wang et al. [71].*

In comparisons between fixed workloads with constant global batch sizes in our data (i.e., LLama-70B 1 vs 8, and Resnet 1 vs 8), we find that increasing the allocation of computing devices (nodes) increases the total IT energy of the workload – increasing by 60.17% and 148.34% for the LLama-70B and ResNet training jobs, respectively. Although more devices increases the speed at which a workload is executed, per-node power draw does not decrease enough to offset the energy cost of additional devices. Practitioners therefore face a three-way tradeoff between training speed, energy consumption, and infrastructure costs that should inform their hardware provisioning decisions [68].

# Appendix B: Inference Energy Estimation

In contrast with training, inference energy consumption has only recently garnered significant scholarly attention [24]. A review of 98 papers modeling AI energy use between 2015 and 2022 found only 17 specifically modeled a distinct inferencing energy footprint [107], though inference computational efficiency has long been a topic of interest among ML researchers. Training has a distinct hardware profile and discrete periods of operation, making it conceptually straightforward[6] to estimate energy footprint [43]. In contrast, inference is a much less distinct load, with more ambiguous hardware. Historic estimates of the inferencing share of ML related energy use for a given AI model deployments range from 35% [33] to 60% [32].

Note that these comparisons of training and inference energy are for historical data before widespread consumer adoption of LLMs. While inference of ML models has been a source of revenue for a decade, these algorithmic ad-targeting and content-recommendation workloads were comparatively modest and computationally streamlined compared to frontier models. Forward looking analysis projects an increasing inference energy share as models become more capable. Median case projection by Schneider Electric estimated inference to draw an 85% share of AI related compute load by 2028 [108].

A major challenge in modeling inference energy use is the heterogeneity of inference hardware. Depending on the output and DNN architecture, ASICs [109], FPGAs [110], GPUs [111], or even CPUs [24], [112] can achieve the highest performance. This then makes modeling of widespread inferencing a challenge, as outside the laboratory inference occurs on a mixed hardware stack. Some research adopts a top-down approach, assuming a given hardware mix, and multiplying number of inference tokens by an efficiency derived from TDP [111] or parameterizing an inference power draw to a fraction of TDP [85], [92]. Modeling LLM inference demand to match search engine request rate, Chien and coauthors estimated inference energy use to be 1,386 times the training cost of GPT-3 [92].

Empirical measurements of hardware power draw are more common in the inference related research than in training [22], [24], [113]. This in part stems from differences in the upfront capital cost of inferencing hardware infrastructure as compared to training. While an individual training node costs on the order of hundreds of thousands of U.S. dollars, inference can be performed in a laboratory setting on commercially available hardware. Luccioni, Jernite, and Strubell used empirical inference performance to estimate the required total queries to match training energy use of differently sized models [22]. They estimated query totals on the order of hundreds of millions to match training costs, with widespread consumer adoption such query totals would occur in a few months. Energy performance of inferencing can also be improved through architectural and algorithmic optimization [25], and carbon aware scheduling [92]. The inference application modeled by Chien and coauthors is consumer facing generating text, a distinctive workload with a unique latency sensitivity. Anecdotal reports of LLM chatbot product development reference introducing additional delay in response display to better match consumer preferences. Other inference loads such as robotics, algorithmic finance, or vehicle automation may be more latency sensitive and thus possess different load shifting feasibility.

---

[6] Note that cloud service providers vary in the level of hardware transparency they afford users, and cloud operators have limited visibility into the algorithms being run on their systems. This informational siloing limits opportunity for hardware-software cooptimization

# Appendix C: Alternative Model Specifications and Component Power Behavior

While our chosen model specification maps well onto our quantity of concern, there may be superior alternatives for physically describing the underlying computational system. One such alternative would be a more robust decomposition of the component contributions to node and system power draw to the component level. These cannot be determined from node level data, but we present our hypotheses for the functional forms relating power draw of each to computational demand below:

- Logic components (G, L): Logic component power consumption follows an asymptotic relationship due to architectural limits in parallelization and fundamental voltage-frequency constraints. At low utilization, power consumption scales roughly linearly with frequency while voltage remains at minimum operational levels, reflecting the sum of constant leakage power and dynamic switching losses. However, as computational demands increase beyond what can be achieved at minimum voltage, higher operating frequencies require increased supply voltages, leading to cubic scaling in power consumption due to the combined effect on both switching and leakage losses. This creates a distinct inflection point in the power-performance curve where the scaling behavior transitions from linear to cubic. GPUs display particularly complex behavior, showing non-linear decreases in power consumption with reduction in arithmetic intensity, suggesting an exotic functional form [49].

- Fans (F): A bank of *n* single speed fans draws power according to stepwise scaling with total heat generation until all *n* fans are in use, which represents a ceiling of power demand. Alternatively, if each fan is spun by a driven motor capable of variable speed operation, there will be a more complicated scaling as each individual fan draws power depending on the cube of its speed [114], [115]. Fan controllers could then optimize the energy performance of the fan bank depending on the scale and distribution of the thermal load. Precisely characterizing these dependencies, and the energy performance impact from fully shifting thermal management of AI training servers to more efficient technologies, is an opportunity for future empirical work.

- Memory and storage systems (S): High-bandwidth memory (HBM) power demand increases step-wise linear as memory banks are activated, up to maximum capacity. Storage drive energy use during batch loading and unloading should linearly scale with model size up to a maximum. Standalone metering of the power draw of integrated memory components (as opposed to storage) is an open technical problem, and research should prioritize a non-invasive solution which could generalize across hardware configurations.

- Interconnect (I): power consumption likely follows a piecewise linear relationship with data movement demands, characterized by distinct behaviors at different scales. Within nodes, high-speed GPU-to-GPU links (like NVLink) should show approximately linear power scaling with bandwidth utilization until hitting physical channel limits. For multi-node training, the relationship becomes more complex as power consumption depends on both bandwidth utilization and network topology. For multi-node training, both intra-node (between GPUs) and inter-node communication should be metered to better capture total system energy use.

# Appendix D: LOOCV Coefficient, Workload Hyperparameter, and Validation Set Summary Tables

**Table D1: FLOP Calculation Summary by Workload**

| WORKLOAD | ARCHITECTURE | BATCH SIZE | HARDWARE | FLOP CALCULATION METHOD | KEY PARAMETERS |
|---|---|---|---|---|---|
| BNL - RESNET 1 (1) | CNN (ResNet152) | 512 (64/GPU) | 1 node × 8 H100 | 11.3 GFLOPs/image × 512 × 3 | Image size: 32×32, Dataset: CIFAR-10 (50K), Epochs: 200, Learning rate: 0.05 |
| BNL - RESNET 2 (1) | CNN (ResNet152) | 4096 (512/GPU) | 1 node × 8 H100 | 11.3 GFLOPs/image × 4096 × 3 | Image size: 32×32, Dataset: CIFAR-10 (50K), Epochs: 200, Learning rate: 0.05 |
| SMC - RESNET 2 (1) | CNN (ResNet50) | 3200 (400/GPU) | 1 node × 8 H100 | 3.6 GFLOPs/image × 3200 × 3 | Image size: 224×224, Dataset: ImageNet (1.28M), Epochs: 35, Learning rate: 0.011 |
| SMC - RESNET 1 (8) | CNN (ResNet50) | 3200 (50/GPU) | 8 nodes × 8 H100 | 3.6 GFLOPs/image × 3200 × 3 | Image size: 224×224, Dataset: ImageNet (1.28M), Epochs: 35, Learning rate: 0.011 |
| BNL-LLAMA-13B (1) | LLM (Llama2-13B) | 64 | 1 node × 8 H100 | Narayanan formula (Equation 10) | Hidden size: 5120, Layers: 40, Sequence length: 4096, TP: 4, PP: 1, CP: 1, Learning rate: 0.00002 |
| SMC - LLAMA-70B (1) | LLM (Llama2-70B) | 4 | 1 node × 8 H100 | Narayanan formula (Equation 10) | Hidden size: 8192, Layers: 80, Sequence length: 4096, TP: 4, PP: 1, CP: 1, Learning rate: 0.0004 |
| SMC - LLAMA-70B (8) | LLM (Llama2-70B) | 8 | 8 nodes × 8 H100 | Narayanan formula (Equation 10) | Hidden size: 8192, Layers: 80, Sequence length: 4096, TP: 4, PP: 1, CP: 2, Learning rate: 0.00036 |
| SMC - LLAMA-70B (64) | LLM (Llama2-70B) | 64 | 64 nodes × 8 H100 | Narayanan formula (Equation 10) | Hidden size: 8192, Layers: 80, Sequence length: 4096, TP: 4, PP: 1, CP: 2, Learning rate: 0.0005 |
| SMC - GPT3-175B (64) | LLM (GPT3-175B) | 128 | 64 nodes × 8 H100 | Narayanan formula (Equation 10) | Hidden size: 12288, Layers: 96, Sequence length: 2048, TP: 4, PP: 8, CP: 1, Learning rate: 0.00002 |

**Table D2: Full LOOCV Steepness Coefficient Stability**

| LOOCV Results | Asymptotic Model | Sigmoid Model | |
|---|---|---|---|
| Held-out workload | α | k | $x_o$ |
| Full Data | 5.55 | 1.12 | 11.46 |
| SMC - GPT3-175B (64) | 5.98 | 1.28 | 11.45 |
| SMC - Llama-70B (64) | 5.2 | 1.07 | 11.58 |
| SMC - Llama-70B (8) | 6.1 | 1.18 | 11.41 |
| SMC - Llama-70B (1) | 5.58 | 0.97 | 11.46 |
| BNL-Llama-13B (1) | 6.05 | 1.31 | 11.45 |
| SMC - ResNet 2 (1) | 5.7 | 1.11 | 11.45 |
| SMC - ResNet 1 (8) | 5.54 | 0.83 | 11.44 |
| **BNL - Resnet 1 (1)** | **4.64** | **0.006** | **9.21** |
| BNL - Resnet 2 (1) | 5.23 | 1.24 | 11.45 |
| LOOCV Summary | | | |
| | α | k | $x_o$ |
| Mean | 5.56 | 1.01 | 11.24 |
| StdDev | 0.43 | 0.4 | 0.72 |
| CoV (%) | 7.7 | 39.6 | 6.4 |

**Table D3: Validation Dataset Summary**

| Workload | Arch. | Parameters | Nodes | $P_{avg}$ (kW) | Energy (kWh) | Duration (Hours) | FLOPS/Node | Source |
|---|---|---|---|---|---|---|---|---|
| Dell Llama-70B (8) | LLM | 70B | 8 | 7.29 | 5.54 | 0.10 | $2.29 \times 10^{15}$ | Dell |
| SMC Llama-70B (8) | LLM | 70B | 8 | 7.73 | 5.43 | 0.09 | $2.29 \times 10^{15}$ | SMC |
| UNet-19M-1 (1) | CNN | 19M | 1 | 6.41 | 1.43 | 0.22 | $4.87 \times 10^{12}$ | SMC |
| UNet-19M-2 (9) | CNN | 19M | 9 | 5.90 | 3.06 | 0.06 | $6.96 \times 10^{11}$ | SMC |